\documentclass[twocolumn,preprintnumbers,amsmath,amssymb]{revtex4}
\usepackage{graphicx}
\usepackage{dcolumn}
\usepackage{bm}

\begin{document}

\title{ Transverse-Longitudinal Emittance Transfer in Circular Accelerators Revised}
\author{Lajos Bojt\'ar}
\email{ Lajos.Bojtar@cern.ch }
\affiliation{ CERN, CH-1211 Gen\`eve 23\\}
\date{\today}

\begin{abstract}
We revise a previous version of this paper \cite{NonHamiltonian} proposing a method to transfer emittances from the transverse planes to the longitudinal. It turned out the system showed non-Hamiltonian behavior because the longitudinal component of the transverse electric field at ends of the transverse resonator were not included in the model. Including those fields the 6D emittance is conserved. We review the mechanism behind the emittance conservation. It is still possible to  produce an emittance transfer with the RF quadrupole field setting the tunes on synchro-betatron resonance for a short period of time then reducing the transverse RF field to zero. We found some coupling between the transverse and the longitudinal planes is always present, even  there in no special device in the accelerator to provoke it. The role of the transverse RF field is to modify the transverse tunes quickly putting the system to  synchro-betatron resonance for a short time. The emittance tends to flow from the plane with bigger emittance to the smaller one. 

\end{abstract}

\maketitle

\section{ Introduction }
In accelerators one of the most important parameter of a particle beam is its emittance, in many cases reducing the emittance is highly important. The common method to achieve small emittance is beam cooling.  Cooling methods reduce the volume of the 6 dimensional phase-space. There are several types of cooling methods, each having its limitations, and in many cases non of those method can be used, or they are not practical. In certain cases it is more important to have a small emittance in one or two planes of motion than the others. Currently known methods ~\cite{PhysRevSTAB.5.084001},  ~\cite{PhysRevSTAB.9.100702}  for complete emittance exchange between subspaces are subjects to certain restrictions. In  accelerators where the beam transfer is symplectic and linear there are constraints which prevents the continuous transfer of emittances between different planes of motion, only a complete exchange is possible.  This was first proved by E.D Courant ~\cite{Courant} and it is known as the emittance exchange theorem  ~\cite{Kim:2006}. An other method ~\cite{Okamoto} uses coupling resonances in a special ring to change the projections to subspaces. Methods using coupling resonances can produce partial emittance transfers between the planes of motion.  In the previous version of this paper \cite{NonHamiltonian}  we presented a new method believed to produce non-Hamiltonian effects. Unfortunately the non-Hamiltonian behavior was a result of a missing longitudinal component of the transverse electric field at the ends of the resonator which wasn't included in the model. We re-cite the idea described in the previous version  \cite{NonHamiltonian}  and discuss it with the correct model including the end fields. From time to time attempts appear to reduce the 6D emittance by external EM fields.  K.W.Robinson has proved in  the appendix of his paper ~\cite{Robinson}, that the synchrotron  radiation damping partition numbers can not be changed by a general EM field. One aim of this revision is to recall this to who might try similar methods. Another reason  is that it is possible to produce a transverse-longitudinal emittance  transfer with the device described below, although this emittance transfer is subject to all restrictions imposed by symplectic conditions.
An important difference compared to the results obtained in the previous version of this paper  \cite{NonHamiltonian} is that the emittance tends to flow from the plane having the larger emittance to the plane with smaller emittance. Existing proposals  ~\cite{PhysRevSTAB.5.084001},  ~\cite{PhysRevSTAB.9.100702}, ~\cite{Okamoto} focus on application for free electron lasers. Our method doesn't require a separate ring for the emittance transfer, the transverse resonator can be placed in an existing ring. Its role is not to introduce a coupling between the longitudinal and transverse planes, but to bring the transverse tunes to synchro-betatron resonance conditions and moving away quickly when the required amount of transfer has happened. The simulations showed there is always some coupling between the longitudinal and transverse planes, even when the transverse resonator is not present and the accelerating cavity is placed in a dispersion free region. 
\section{ The basic idea revised} \label{basic_idea}
We re-cite the idea presented in the previous version of this paper  \cite{NonHamiltonian}  and also the reason why it was believed to produce non-Hamiltonian effects. Then we describe the mechanism preventing us from using the transverse resonator that way.
Let's consider a particle circulating in a storage ring, that contains at least one accelerating cavity with some RF voltage, and another RF resonator that generates  a transverse quadrupole electric filed.  This second electric field can be given as
  \begin{equation} \label{eq_efield} \textbf{E}\left( x,y,t\right) = E_{g}\left( \textbf{i}\,x -\textbf{j}\,y \right) \cos\left( \omega t+\phi \right)\;\;,  \end{equation}  
where $ E_{g} = \partial \textbf{E} / \partial x$  is the gradient of the electric field in $ \left[ V/m^2 \right] $ and \textbf{i, j} are the Cartesian unit vectors. 
The electric field in  eq. ~(\ref{eq_efield}) can be derived from the vector potential, the magnetic field obtained from the same vector potential is zero.
 \begin{eqnarray} \label{vector_pot}
 \textbf{A}\left( x,y,t\right) &=& -A_{g}\left( \textbf{i}\,x -\textbf{j}\,y \right) \sin\left( \omega t+\phi \right)\;\;,  \\
 \textbf{B} &=&  \nabla  \times \textbf{A} =0 \;\;\;, \\
 \textbf{E} &=& -\frac{d\textbf{A}}{dt}\;\; .
 \end{eqnarray} 
The equations above give a good model of the fields inside the transverse resonator (TR from here). There are however longitudinal components of the electric field at the ends of the TR  not included in the model of our first attempt. These end fields can be modeled as a longitudinal kick giving an energy change to the particle
\begin{equation}\label{eq_end_field}
\Delta E =\pm \frac{1}{2} (x^2-y^2) \cos( \omega t+\phi )\;\;.
 \end{equation}  
It is straightforward to obtain eq.(~\ref{eq_end_field}). If we choose the potential to be zero outside the TR, a particle traveling through the end field will get a longitudinal kick corresponding the the potential difference, which is proportional to the square of its transverse position. The directions of the end fields are opposite at the entry and the exit.

Lets consider first the effect of the transverse electric field inside the TR. The direction of the field points inward in the horizontal plane and points outward in the vertical plane. Fig.~\ref{fig:assembly} and Fig.~\ref{fig:efield} might help visualize this field.  Imagine a beam  diverging in the horizontal plane and converging in the vertical. A positively charged particle traveling in such a beam looses kinetic energy through the RF electric field, because its motion most of the time has transverse components which are opposite to the direction of the force acting on it. There is an energy flow outward from the system each time a bunch of particles goes through the TR. The energy loss due to the transverse electric field is compensated to some extent by the energy change due to the end fields. At the electrostatic limit the energy changes due to the transverse and longitudinal components cancel each other exactly. There is a frequency and length of the TR at which a the phase of the RF field changes $\pi$ radians while a  synchronous particle travels through the TR. In this case the energy change is purely due to the transverse electric field. The amount of the energy loss depends on the angle of the particle's trajectory and also on its longitudinal position. In the accelerating cavity the particle gains or loses energy depending only on its phase with respect to the synchronous particle.  This later energy change is strictly a function of phase, while the energy change in the TR is a function of the transverse trajectory inside that field and also the phase of the particle with respect to the transverse RF field. It is not easy to see why the energy flowing outward the system at the TR equals exactly the energy change in the accelerating cavity averaged over some period of time.

The Panofsky-Wenzel theorem ~\cite{panofsky:967} already indicated why an EM field can not reduce the 6D emittance. Later Hereward \cite{Hereward} found that for any electromagnetic field in a cavity the following equation must hold as a consequence of Maxwell's equations.
\begin{equation}\label{Herewards_eq}
\frac{\partial dp_{x}}{\partial z} = \frac{\partial dp_{z}}{\partial x}\;\; ,
 \end{equation}  
where z is the longitudinal position deviation from the synchronous particle. Eq. ~(\ref{Herewards_eq}) shows that the force acting on the particle is a conservative one. In the appendix of \cite{Robinson} Robinson obtained the same equation and also gave a detailed derivation of why a general EM field can't change the synchrotron radiation damping partitions. The same mechanism is valid for our case. We re-cite here only  the main points from \cite{Robinson}. The equation of the longitudinal motion ignoring the synchrotron radiation can be written as
\begin{eqnarray} \label{synchrotron_motion_eq}
\frac{dE}{dN} &= \frac{\partial \epsilon}{\partial z} z +\frac{\partial \epsilon}{\partial E} \Delta E \;\; , \nonumber \\
\frac{dz}{dN} &= -\frac{\partial l}{\partial E} \Delta E  -\frac{\partial l}{\partial z} z \;\; .
\end{eqnarray} 
Where $\epsilon$ is the energy change per turn due to the EM fields in the ring, $l$ is the path length, E is the energy of the particle and $z$ is the longitudinal position deviation from the synchronous particle. There are two terms in the equations above which can produce damping and it is shown in \cite{Robinson} that these two terms cancel each other as a consequence of eq.(\ref{Herewards_eq})
\begin{equation}\label{damping_terms}
\frac{\partial \epsilon}{\partial E} = \frac{\partial l}{\partial z}\;\; .
 \end{equation}  
In other words, the longitudinal damping produced by the energy loss depending on the energy is compensated by the anti damping due to the path length dependence on longitudinal position or vice versa. The path length depends on the longitudinal position because it is proportional to the  transverse deflection introduced by the TR field, which is itself dependent on the longitudinal position of the particle. 

An other similar case has been analyzed in ~\cite{Cornacchia:282603}, where a dipole mode cavity was placed in a dispersive region. They came to the same conclusion, energy loss dependending on the energy is compensated by the the path length change introduced by the dipole mode cavity.

In both case  using the TR or a dipole mode cavity, the force introduced by the EM field is a conservative, therefore  Liouville's theorem applies. The 6D emittance stays constant. Even with this constraint it is possible to introduce correlations between the planes of motion and change the projected emittances using synchro-betatron resonances.

\section{ Transverse electric field generation \protect\\ }

\begin{figure}
\includegraphics [scale=0.5]{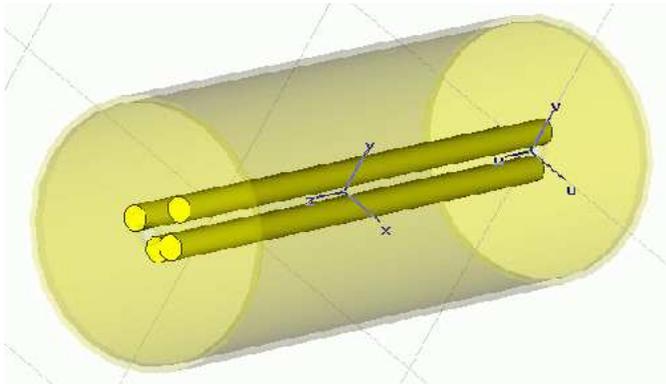}
\caption{\label{fig:assembly} One possible design of a TR to generate quadrupole electric field}
\end{figure}

\begin{figure}
\includegraphics  [scale=0.6] {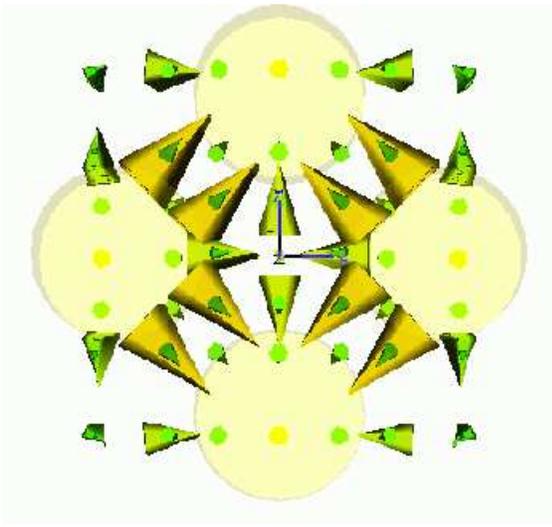}
\caption{\label{fig:efield} The electric field between the the rods.}
\end{figure}

It isn't an aim of this paper to present a detailed design of the apparatus, but to demonstrate that the method is feasible and the electric field  of the TR can be practically realized. The electric field is similar to the field of  an RFQ, but the rods have no undulations. The frequency of the TR must be relatively low, because  during the time a bunch travels through the TR the RF voltage should not change sign. One possible structure is described in ~\cite{rfq_patent}. This type of design has the advantage of having a resonant frequency inversely proportional to its length. The RFQ   in ~\cite{rfq_patent}  without vane modulation has been modeled with Microwave Studio  ~\cite{cst_ms}. The structure is shown on Fig.~\ref{fig:assembly}. The diameter of the cylinder is 1 m, it's length is 2.5 m, the diameter of the rods are 8.72 cm. The electric field between the rods has the best linearity if the ratio between the radius of the rods and their distance from the center is 1.14511. With these dimensions the structure resonates at 13.56 MHz. The frequency can be lowered making the structure longer. Fig.~\ref{fig:efield}  shows the electric field from a view point on the beam axis. The magnetic field is mainly located outside the rods and has little effect on the beam.

\section{ Simulations\protect\\ }

\subsection{ The method}

A Java code has been developed from scratch for the simulations. It was believed the TR produce non-Hamiltonian effects, and that was the main reason to develop the code and not using standard tools. The effect we are interested in  accumulates over thousands of turns, so it was important to avoid any approximations and have a full understanding of what the code does. Classical methods like Runge-Kutta-Cash-Karp with automatic step size adjustment was found to be precise enough for the number of turns simulated. The source code is freely available from the author on request.  Five type of elements have been implemented: Sector bending, magnetic quadrupole, RF accelerating cavity, drift space and the TR. Simulations used some combination of these elements. All elements were treated with the hard-edge model, expect the TR where the end fields turned out to be essential. The numerical precision has been verified by performing  time reversed tracking. The initial conditions of the forward tracking has been  re-obtained with 6-7 digit precision after time reversed tracking (20000 traversals of the FODO cell).

\subsubsection{ Sector bending magnet}

The trajectory in the sector bending magnet is treated in a geometric way in a global coordinate system, where the $x,y$ plane is the bending plane and the  $ z $ coordinate is the vertical one. The particle arriving in the bending magnet moves on a circle with a radius determined by 
\begin{equation}
\rho = \frac{p_{xy}}{B q}\;\;,
\end{equation}
 where 
\begin{equation}
p_{xy} =\gamma m_{0} \sqrt{v_{x}^2+v_{y}^2}\;\;.
\end{equation}
The center of the circle lies on the line perpendicular to the particle's trajectory at the entry. Then the intersection of the trajectory circle and the exit edge is calculated, that gives the $x$ and $y$ coordinates at the exit. The $v_{x}$,$v_{y}$ velocity components are rotated in a global coordinate system by the angle extent of the arc between the entry and the exit points, $v_{z} $ is unchanged . The $z$ coordinate at the exit is calculated as $z=z_{0}+v_{z} L/v_{xy}$, where $L$ is arc length between the entry and the exit points.

\subsubsection{ Quadrupole}

The global coordinates and velocities are transformed to a local one, then the  trajectory is calculated by solving the equations of motion numerically, and finally the local coordinates and velocities are transformed back to the global coordinate system. In the local coordinate system $x$ stands for the horizontal position $y$ for the vertical and $z$ for the longitudinal. The equation of motion is obtained as
\begin{eqnarray}
\frac{d\textbf{p}}{dt} &=& q(\textbf{B}\times\textbf{ v})\;\;, \nonumber \\
\textbf{B}  &=& \lbrace B_{g} x\left( t \right) , B_{g} y\left( t \right) ,0  \rbrace \;\;, \nonumber \\
B_{g}&=& \frac{B_{x}}{dx}\;\;,
\end{eqnarray}
which gives
 \begin{eqnarray}  
    x''(t) +K_{m} x z'(t) = 0\;\;, \nonumber\\
    y''(t) - K_{m} y z'(t) = 0\;\;,  \nonumber\\
    z''(t) - K_{m}  \left( x x'(t) -y y'(t) \right)  = 0\;\;,
\end{eqnarray} 
with
\begin{equation}
 K_{m}= \frac{ B_{g} q }{m_{0} \gamma}\;\; . 
\end{equation}

Please note that there is a  longitudinal component of the force, which is missing in the usual approximative formalism. The independent variable is time. The precise time when the particle exits the quadrupole isn't known in advance, it is found by successive approximations.

\subsubsection{ Cavity}

The longitudinal component of the momentum is increased in a single step  at the middle of the cavity's physical length, by the amount which  corresponds to the energy change: $ \Delta E =V_{0} \sin(\omega t +\phi) $ in $ \left[ eV \right].$ The spaces before and after the middle point are treated as drift space. The transverse velocities are modified to keep the transverse momentum unchanged.

\subsubsection{ Transverse quadrupole  resonator}

The TR field is handled in two parts. Longitudinal components  at the ends are modeled as a longitudinal kick. The transverse  component inside the TR are modeled similarly to the magnetic quadrupole. The equation of motion is derived starting with $d\textbf{p}/dt =  \textbf{E} q$  using  eq.~(\ref{eq_efield})  and taking into account that the $ \gamma $ is changing during the transfer, we obtain
\begin{eqnarray}
x''(t) \gamma (t)  + x'(t) \xi (t)  -\frac{E_{g} q}{m_{0} } \cos (\phi + \omega t ) x(t) =0\;\;, \nonumber\\
y''(t) \gamma (t)  + y'(t) \xi (t) +\frac{E_{g} q}{m_{0} } \cos (\phi + \omega t ) y(t) =0\;\;, \nonumber\\
z''(t) \gamma (t)  + z'(t) \xi (t)    =0\;\;, \nonumber\\
\end{eqnarray}

where 
\begin{equation}
\xi(t) =\frac{  \gamma(t)^3  \left(x'(t) x''(t)+ y'(t) y''(t)+ z'(t) z''(t)\right)} { c^2}\;\;. 
\end{equation}
The same equation was derived from the relativistic Lagrangian as a cross check. 
We ignored the magnetic field inside the rods in this model. There is some weak magnetic field between the rods, but if we ignore both this weak magnetic field and the small longitudinal variation of the transverse electric field our model stays consistent. 

At the entry and the exit of the TR there are longitudinal components of the EM field which must be taken into account. These end fields are modeled as longitudinal kicks. The procedure is very similar to the one used for the accelerating cavity with some difference. The energy change is not only a function of the particle's phase respect to the synchronous particle, but also a function of the transverse position at the entry and at the exit. The energy change is given by eq.(\ref{eq_end_field}).

\subsection{ Tracking  results}
As we have already seen, the force introduced by the TR can not produce damping. When the  end fields of the TR  were included, the damping of the betatron oscillations and the increase of the synchrotron oscillation amplitude described in the first attempt of this paper has disappeared. The transverse and longitudinal emittances stayed constant. There are however special conditions when the emittance can be transferred between the planes. When one of the transverse tune and the synchrotron tune satisfy the condition
\begin{equation}\label{condition}
n q_{\bot} = m q_{s} 
\end{equation}
\begin{figure}
  \begin{center}
    \begin{tabular}{c}
      \resizebox{80mm}{!}{\includegraphics{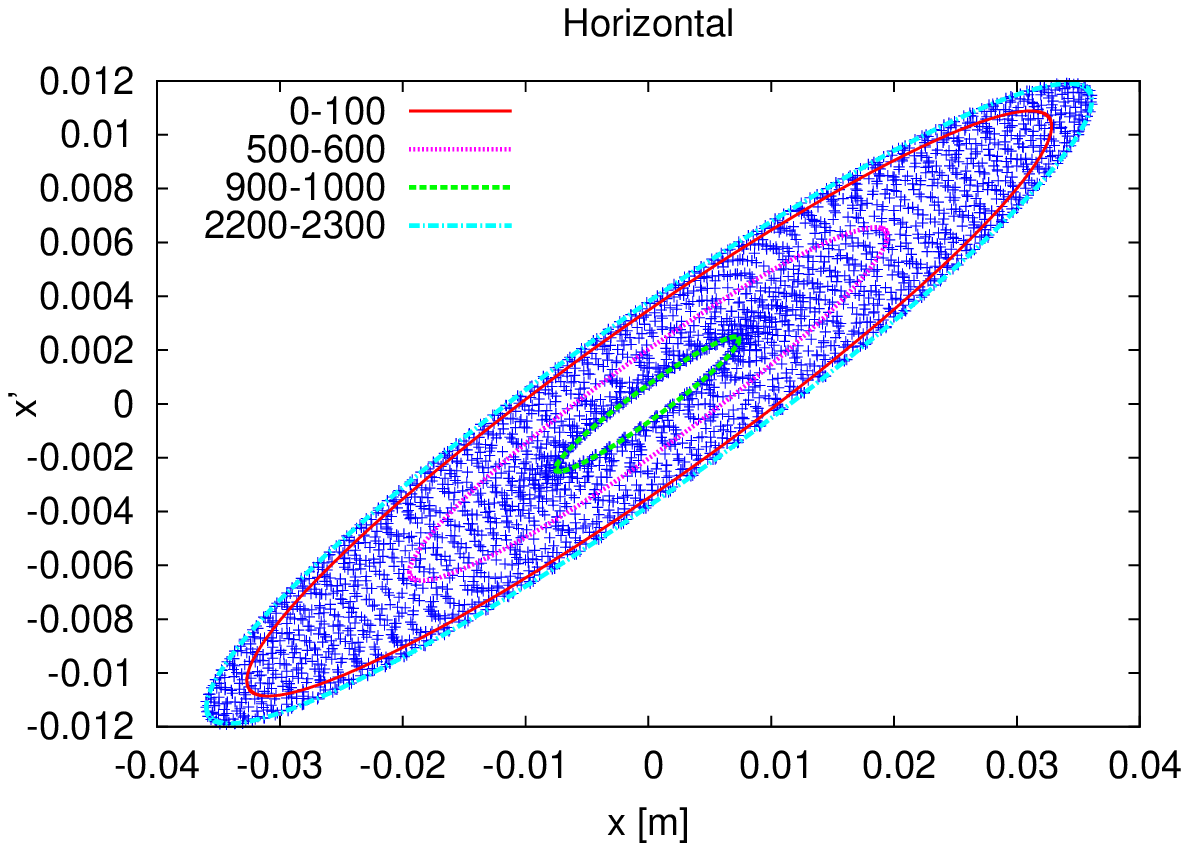}} \\
      \resizebox{80mm}{!}{\includegraphics{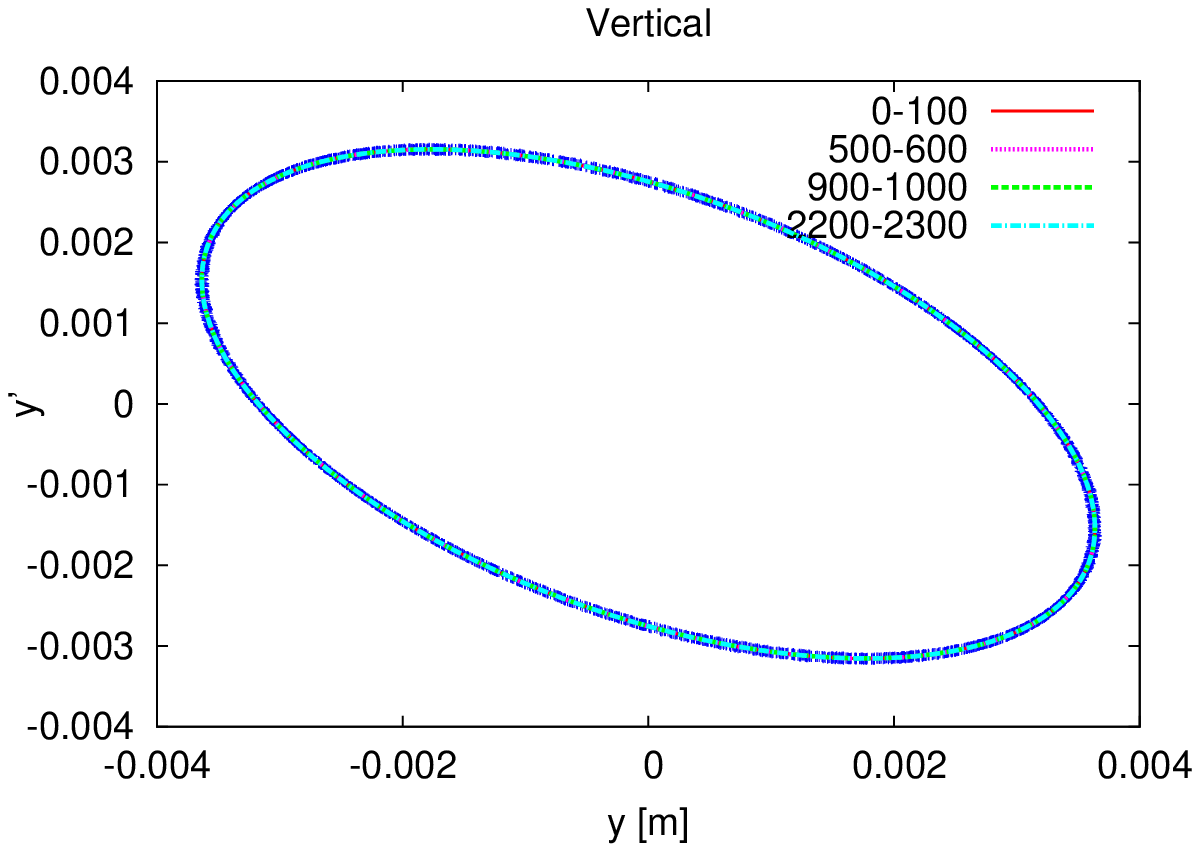}} \\
      \resizebox{80mm}{!}{\includegraphics{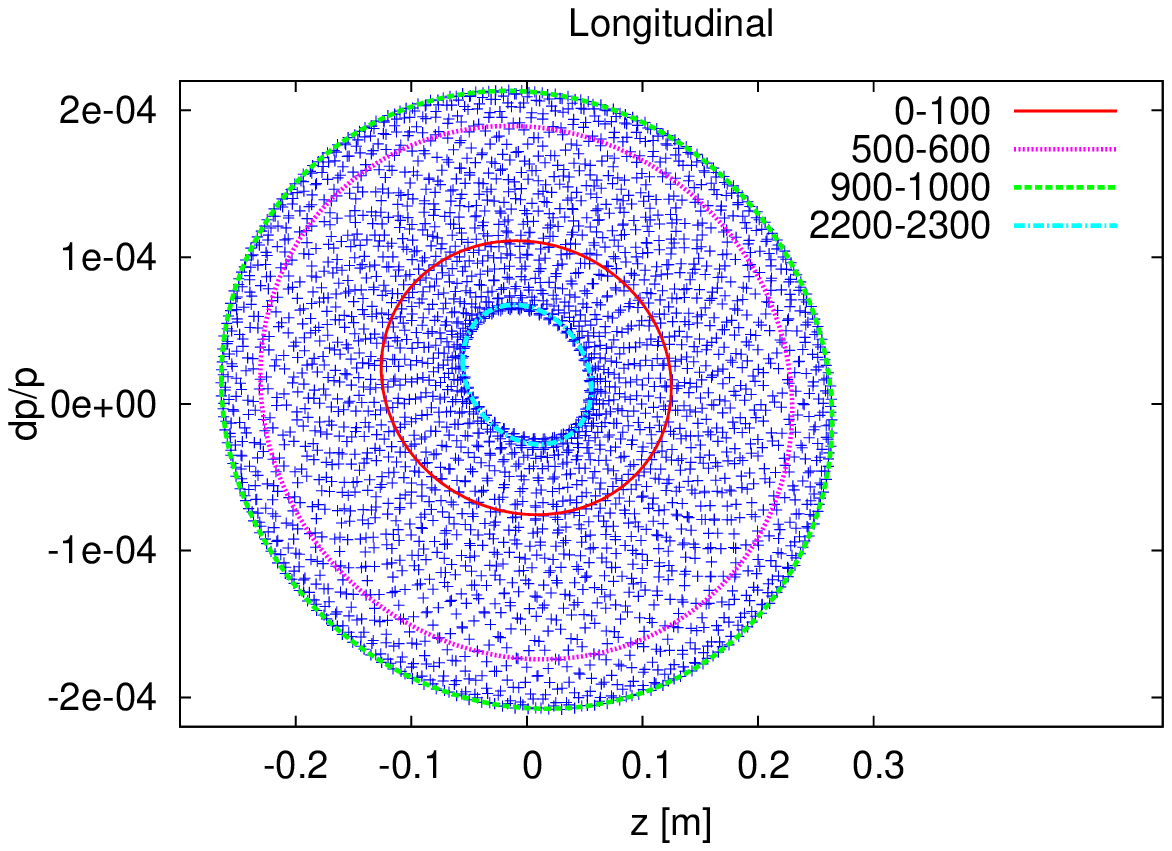}} \\     
    \end{tabular}
  \caption{\label{fig:phs_all}  Single particle phase-space plots showing synchro-betatron resonance between the horizontal and longitudinal planes. The ellipses are fitted to points at different stages during the emittance transfer. } 
  \end{center}
\end{figure}
\begin{figure}
  \begin{center}
    \begin{tabular}{c}
      \resizebox{80mm}{!}{\includegraphics{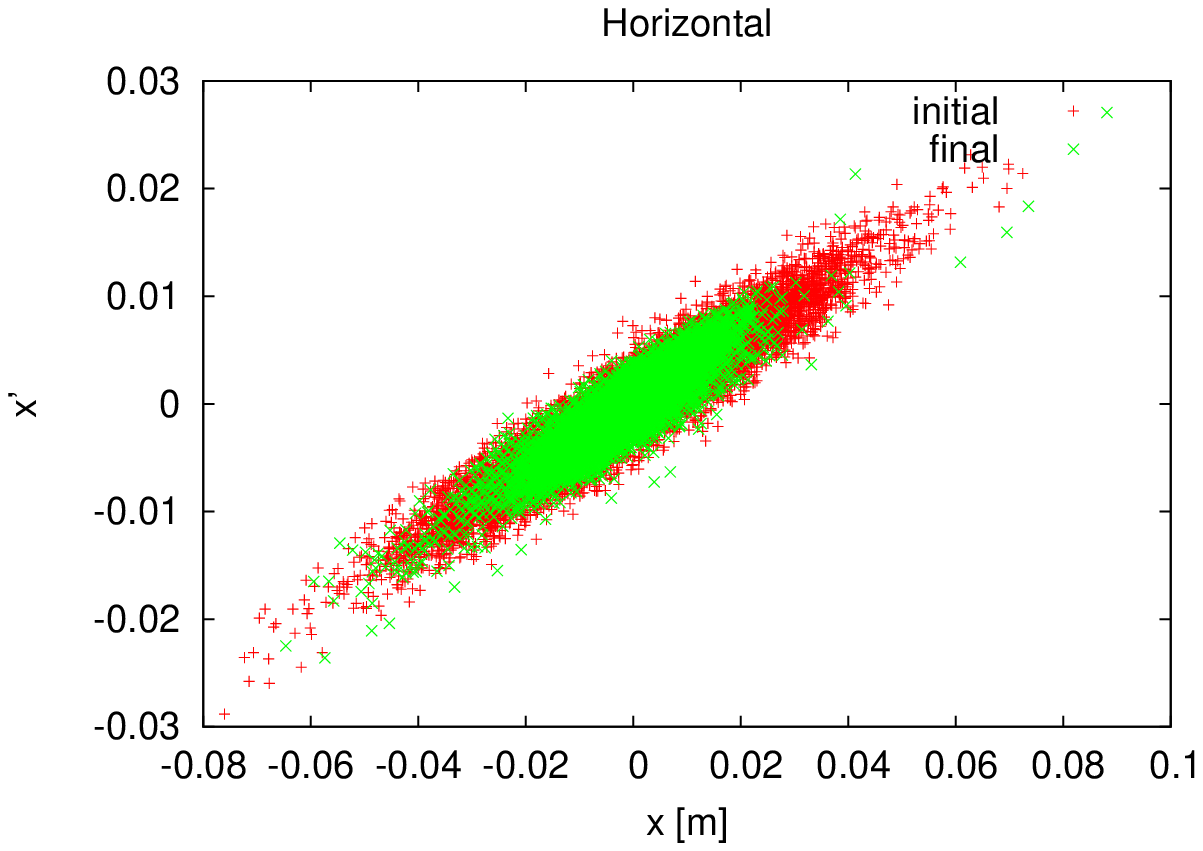}} \\
      \resizebox{80mm}{!}{\includegraphics{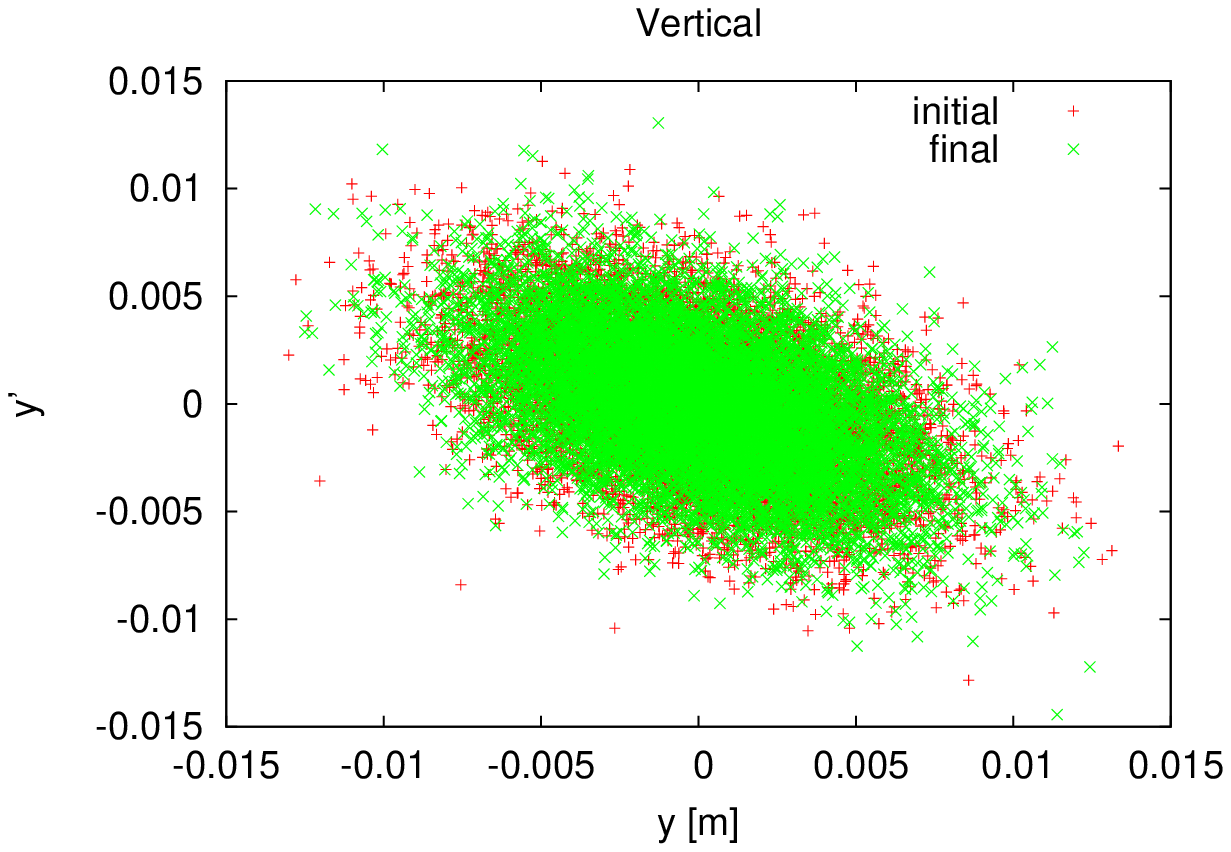}} \\
      \resizebox{80mm}{!}{\includegraphics{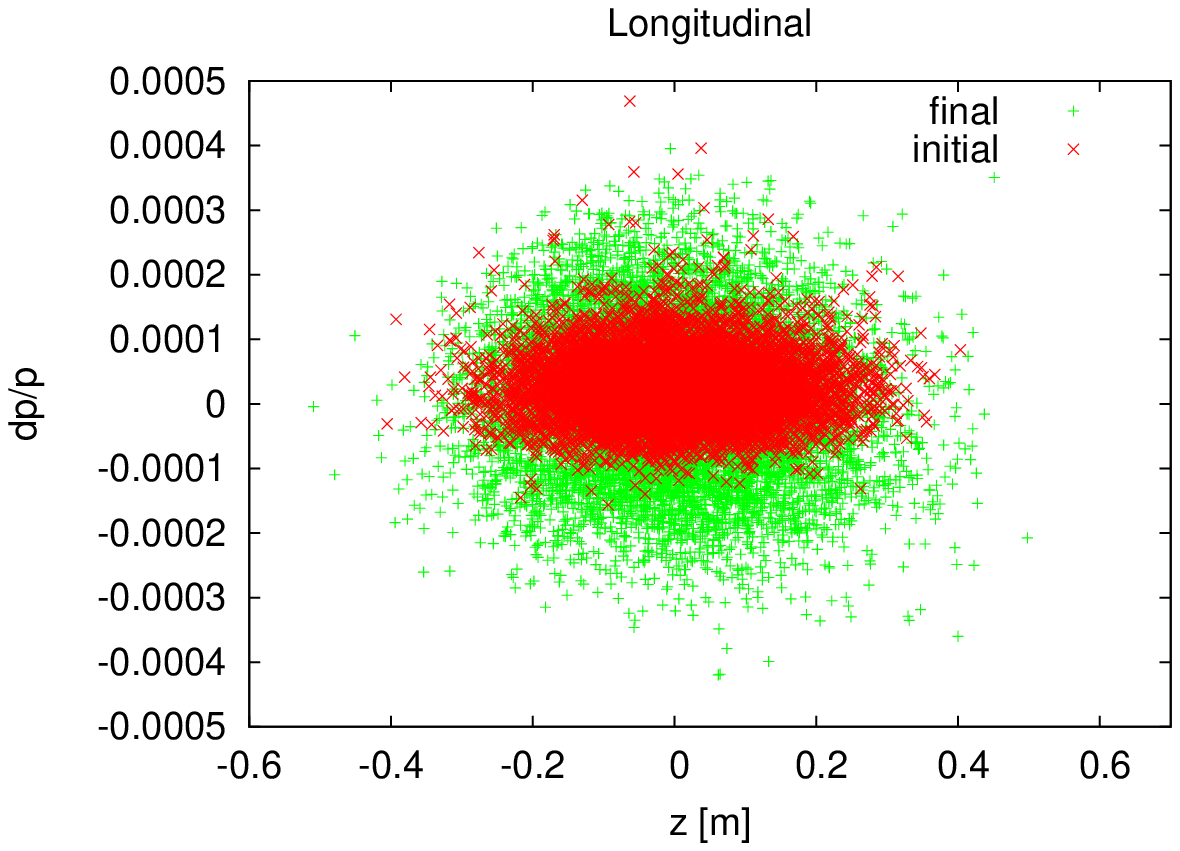}} \\     
    \end{tabular}
    \caption{Initial distributions of 9621 points in phase-space and after 1300 turns.}
    \label{fig:distrib}
  \end{center}
\end{figure}
synchro-betatron resonance occurs. Where n,m are integers and q is the fractional part of the tune. Usually the the coupling is introduced by some device, like a dipole mode cavity in a dispersive region. We found that a transverse-longitudinal coupling is always present without the need for any special device. Even when we simulated the simplest case, a FODO cell consisting of a focusing and a defocusing magnetic quadrupole and an accelerating cavity, the synchro-betatron resonance was present, if the tunes were adjusted to satisfy the resonance condition in eq.(\ref{condition}). The path length of a particle always depends to some extent on its betatron motion.

 A particle with small betatron amplitude has a shorter path length than a particle with a large betatron amplitude.
We made simulations with machines containing  the TR, then the TR was replaced by a magnetic quadrupole giving the same focusing strength. There was no noticeable difference between the behavior of the two system. If the TR introduce some coupling between the transverse and longitudinal planes it is much smaller than the intrinsic coupling, which is present even without it. What is the role of the TR after all? The emittance tends to flow from the plane with bigger emittance to the the smaller. This process has to be stopped at a certain moment to be useful. In ~\cite{Okamoto} a dedicated ring is proposed for the emittance transfer tuned on synchro-betatron resonance. The beam is injected into that ring, then it makes a few hundred turns while the emittance transfer is taking place, then it is ejected.
The TR can be useful for emittance transfer in existing rings. Its role is to modify the transverse tunes rapidly to satisfy the condition in eq.(\ref{condition}), then move the transverse tunes away from the synchro-betatron resonance. With normal magnetic quadrupoles this can not be done, they are too slow. The voltage of the TR can be changed fast enough to do that. There is an other possibility to introduce resonance conditions quickly by changing the voltage of the accelerating cavity. This way the longitudinal tune is changed. This second method has a serious disadvantage. The longitudinal tune is usually a small value, in our simulations it was 0.024. This means the transverse tune has to be very close to an integer value. With a transverse tune so close to integer resonance the beam would suffer losses during a longer period of time. Using the TR the transverse tune can be set to this dangerous value only for the duration of the emittance transfer, in our simulation it was around 1300 turns.

 Fig.(\ref{fig:phs_all}) shows singe particle simulation results in a ring at 1.1 GeV/c. The tunes are $q_h = 7.0244 , q_v = 7.281 , q_l = 0.024$.
These tunes were obtained with four TR in the ring , one in each quadrant giving 11.8 MV/m electric field. These tunes correspond to the case when $ m=n=1$ in eq.(\ref{condition}). The ellipses are fitted to the phase space points at different times. As it can be seen on Fig.(\ref{fig:phs_all}) the emittance start to flow from the bigger to the smaller one.  It is possible to drive the beam from the smaller emittance to the bigger one when $n \neq m$ in eq.(\ref{condition}), but when n or m bigger than 1 the resonance range is very narrow. The transverse tunes have a small amplitude dependence, this leads to  smearing of the phase space as it is noticed in ~\cite{Okamoto}.

When the TRs were off, the horizontal tune was far from resonance, $q_h = 7.1$.  

 Fig.(\ref{fig:distrib}) shows bunch tracking results. Emittances were calculated before and after tracking with the method given by Wolski ~\cite{Wolski}. His method can be used even if there are correlations between the planes . Here is a short summary. The beam $ \Sigma $ matrix is defined as $ \Sigma_{ij} = \langle{( X_{i}-\overline{X_{i}} ) ( X_{j} - \overline{ X_{j} } ) } \rangle $, where $ X^T =(x,x',y,y',z,\frac{\Delta p}{p} ) $ and a matrix S defined as

\begin{equation} \label{sigmam}
S =\begin{pmatrix} 
0  & 1 & 0 & 0 & 0 & 0 \\
-1 & 0 & 0 & 0 & 0 & 0 \\
0 & 0 & 0 & 1 & 0 & 0 \\
0 & 0 & -1 & 0 & 0 & 0 \\
0 & 0 & 0 & 0 & 0 & 1 \\
0 & 0 & 0 & 0 & -1 & 0 \
\end{pmatrix}.  \nonumber\\
\end{equation} 

Then the eigenvalues of $ \Sigma S$ are $\pm  i \varepsilon_x$ , $\pm  i \varepsilon_y$ , $\pm  i \varepsilon_z$.
The brackets $ \langle \rangle $ mean averaging over all particles.
The result of the emittance calculations are in TABLE \ref{table:bunch}. As it can be seen, the 6D rms emittance is slightly bigger after tracking.  The RMS emittance can grow even the system is symplectic. We think it is due to the nonlinear nature of the coupling between the transverse and longitudinal planes. The product of the three eigenvalues, which is the 6D RMS emittance is approximately conserved.
\begin{table}[h]
\centering 
\begin{tabular}{ c c c c c } 
\hline\hline 
  & $ \epsilon_{h} $ &  $\epsilon_{v} $  & $\epsilon_{l}$  & $\epsilon_{6D} $ \\
 [0.5ex] 
\hline 
initial & 39.3 &  9.87  & 5.02 &  1.95E-9  \\
final  & 16.4  & 9.86  & 13.8  & 2.24E-9 \\
[1ex] 
\hline 
\end{tabular}
\caption{ Initial Rms emittances and final emittances after 1300 turns . Emittances ($ 1 \sigma $) are in [mm mrad].}
\label{table:bunch} 
\end{table}

\begin{figure}
\includegraphics  [scale=0.6] {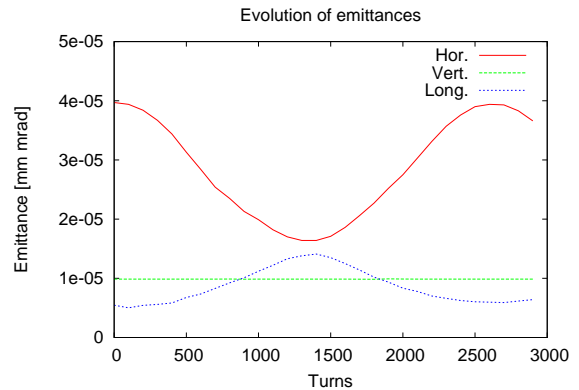}
\caption{\label{evolution}The evolution of emittances as a function of turns.}
\end{figure}

Fig.(\ref{evolution}) shows the evolution of the emittances calculated from the $\Sigma$ matrix as mentioned before. 
The initial distributions are Gaussian and without correlations. About 1300 turns later the horizontal and longitudinal planes have the maximum amount of correlation, at this point the two emittances come closest to each other. An other 1300 turns later the emittances are nearly back to the initial conditions, then the cycle starts again.

\section{ Summary and outlook}
We have revised the idea presented in the first attempt of this paper and  re-cited the mechanism preventing us from getting non-Hamiltonian effects. Maxwell's equations imply eq.(\ref{Herewards_eq}), then it follows that the two damping terms in the equations of synchrotron motion are equal and compensate each other. The force acting on the beam is conservative. We found there is an inherent coupling between the transverse and longitudinal planes. This coupling is present even when there is no any special device in the ring to provoke it. We proposed to use the TR to put the tunes of a ring quickly to synchro-betatron resonance for the duration of the emittance transfer, then  to move the tunes away from resonance. With our method there is no need for a separate ring. It can have applications, where transferring the bigger emittance from one plane towards the plane with smaller emittance has an advantage.

\section{ Acknowledgments}
The previous version of this paper  \cite{NonHamiltonian} aroused vivid discussions. I would like to express my gratitude towards all people who gave their valuable comments and led me to find out the problem with the model. Andrew Sessler, Hiromi Okamoto, Dieter Mohl, John Jowett, Elias Metral, Alessandra Lombardi, Frank Zimmermann, Oliver Bruning, Werner Pirkl. 

I would like to thank Sara Lanzone for setting up PATH simulations and Kevin Priestnall for checking this report gramatically.
\bibliography{PhaseSpaceManipulations} 

\end{document}